\documentclass[a4paper,10pt]{article}
\usepackage[dvips]{graphicx}
\usepackage{amssymb,amsmath}
\numberwithin{equation}{section}

\begin{document}
\title{ Darboux Transformation and Exact Solutions of the Myrzakulov-Lakshmanan-II Equation}
\author{ M. Zhassybayeva, G. Mamyrbekova,  G. Nugmanova,  R. Myrzakulov\footnote{Email: rmyrzakulov@gmail.com} \\ \textit{Eurasian International Center for Theoretical Physics and  Department of General } \\ \textit{ $\&$  Theoretical Physics, Eurasian National University, Astana 010008, Kazakhstan}}


\date{}
\maketitle
\begin{abstract}
The Myrzakulov-Lakshmanan-II (ML-II) equation is one of a (2+1)-dimensional generalizations of the Heisenberg ferromagnetic equation. It is integrable and has a non-isospectral Lax representation. In this paper, the Darboux transformation (DT) for  the ML-II equation  is constructed. Using the DT, the 1-soliton and 2-soliton solutions of the ML-II equation are presented.

\end{abstract}
\vspace{2cm}
\section{Introduction}

During the past decades, there has been an increasing interest in the investigation  of integrable classical and quantum systems.  The theory of  integrable systems is an important branch of nonlinear science. Integrable systems describe various kinds of nonlinear phenomena, such as soliton signals, soliton waves and etc. At the same time, the soliton theory gives many methods of finding exact solutions of nonlinear ordinary  and partial differential equations of modern  mathematical physics.  Constructing exact solutions of such nonlinear differential equations is a hard 
job. However, during the past decades, some methods to find exact solutions are proposed, such as Hirota method, Backlund transformations, Darboux transformations and etc. In this paper, we will construct the Darboux transformation (DT) 
for the Myrzakulov-Lakshmanan-II equation (ML-II equation) and using the DT, some   exact soliton solutions of this equation are found. Note that the DT for some other Heisenberg ferromagnetic equations (HFE) were constructed in \cite{Gut}-\cite{Chen} (see also Refs. \cite{R13}-\cite{M2001}). 

The paper is organized as follows.  In section 2, the ML-II  equation and its Lax representation are introduced.   In section 3, we derived the DT of the ML-II equation.   Using the one-fold DT, some exact  soliton solutions are derived in section 4.  Section 5 is devoted to conclusion.

\section{The Myrzakulov-Lakshmanan-II equation}
Let us we consider the Myrzakulov-Lakshmanan-II equation or shortly ML-II equation. It  looks like \cite{R14}
\begin{eqnarray}
iS_{t}+\frac{1}{2}[S, S_{xy}]+iuS_{x}+\frac{1}{\omega}[S, W]&=&0,\label{2.1}\\
u_x-\frac{i}{4}tr(S[S_x,S_y])&=&0,\label{2.2}\\
 iW_{x}+\omega [S, W]&=&0,\label{2.3}
\end{eqnarray} 
 where  
 \begin{eqnarray}
S=\begin{pmatrix} S_3&S^{-}\\S^{+}& -S_3\end{pmatrix}, \quad W=\begin{pmatrix} W_3&W^{-}\\W^{+}& -W_3\end{pmatrix}.\label{2.4} 
\end{eqnarray} The ML-II equation  is integrable by the IST.  Its   Lax representation reads as \cite{R14}
 \begin{eqnarray}
\Phi_{x}&=&U\Phi,\label{2.5}\\
\Phi_{t}&=&2\lambda\Phi_y+V\Phi,\label{2.6} 
\end{eqnarray}  
where  
 \begin{eqnarray}
U&=&-i\lambda S,\label{2.7}\\
V&=&\lambda V_{1}+\frac{i}{\lambda+\omega}W-\frac{i}{\omega}W\label{2.8} 
\end{eqnarray} 
with\begin{eqnarray}
V_1=2Z=\frac{1}{2}([S, S_{y}]+2iuS).\label{2.9} 
\end{eqnarray}

 \section{Darboux transformation}
 
In this section, our aim is to construct  the DT for the ML-II equation. We first consider the one-fold DT in detail. Then we consider briefly the two-fold  and $n$-fold DT.

\subsection{One-fold DT}
We start from  the following transformation for the any two solutions of the equations (\ref{2.4})-(\ref{2.5}):
\begin{eqnarray}
\Phi'=L\Phi, \label{3.1}
\end{eqnarray}
where we assume that
\begin{eqnarray}
L=\lambda N-K. \label{3.2}
\end{eqnarray}
Here
\begin{eqnarray}
N=\begin{pmatrix} n_{11}&n_{12}\\n_{21}&n_{22}\end{pmatrix}. \label{3.3}
\end{eqnarray}
Of course that the matrix function  $\Phi^{\prime}$ satisfies the same Lax representation as (\ref{2.4})-(\ref{2.5}) so that
\begin{eqnarray}
\Phi'_{x} &=& U'\Phi',\label{3.4}\\
\Phi'_{t} &=& 2\lambda\Phi'_{y}+V'\Phi'. \label{3.5}
\end{eqnarray}
The matrix function $L$ obeys the following equations
\begin{eqnarray}
L_{x}+LU &=& U'L,\label{3.6} \\
L_{t}+LV &=& 2\lambda L_{y}+V'L. \label{3.7}
\end{eqnarray}
From the first equation of this system we obtain
   \begin{eqnarray}
   N&:& \lambda_x=0	\label{3.8}\\
   	\lambda^0&:&  K_{x}=0\label{3.9}\\
	 \lambda^1&:& N_{x}=iS^{'}K-iKS, \label{3.10}\\
 \lambda^2&:& 0=-iS^{'}N+iNS. \label{3.11}
\end{eqnarray}
Hence we have the following DT for the matrix $S$:
\begin{eqnarray}
   S^{'}=NSN^{-1}. \label{3.12}
\end{eqnarray}
At the same time, the second equation of the system  (\ref{3.6})-(\ref{3.7}) gives us 
\begin{eqnarray} 
N&:&\lambda_t=2\lambda\lambda_y,\label{3.13}\\
\lambda^{0}&:& -K_{t}=iW'N+\frac{i}{\omega} W'K-iNW-\frac{i}{\omega}KW, \label{3.14}\\
\lambda^{1}&:& N_{t}=-2K_y-\frac{i}{\omega} W'K-2Z^{'}K+\frac{i}{\omega} NW+2KZ, \label{3.15}\\
\lambda^{2}&:&0=2N_y+2Z^{'}N-2NZ,\label{3.16}\\
(\lambda+\omega)^{-1}&:&0=-i\omega W^{'}N-iW'K+i\omega NW+iKW.\label{3.17} 
\end{eqnarray}
These equations  tell us that
\begin{eqnarray}
   K_{t}=0\label{3.18}
\end{eqnarray}
so that we can put
\begin{eqnarray}
   K=I.\label{3.19}
\end{eqnarray}
Also they give the following DT for the $W$:
\begin{eqnarray}
W'=(I+\omega N)W(I+\omega N)^{-1}. \label{3.20}
\end{eqnarray}
For the unknown matrix function  $N$ we get the equations
\begin{eqnarray}
   N_{x}&=&i(S^{'}-S),  \label{3.21}\\
   N_y&=&-Z^{'}N+NZ,\label{3.22}\\
    N_{t}&=&-\frac{i}{\omega} W'N-2Z^{'}+\frac{i}{\omega} NW+2Z. \label{3.23}
\end{eqnarray}
Hence  we get the second form of the DT for $S$ as:
\begin{eqnarray}
   S^{'}=S-iN_{x},  \label{3.24}
\end{eqnarray}
Also let us calculate $Z^{\prime}$. We have
\begin{eqnarray}
Z^{\prime}=\frac{1}{2}(S^{\prime}S^{\prime}_{y}+iu^{\prime}S^{\prime}).\label{3.25} 
\end{eqnarray}  
After some algebra this equation takes the form
\begin{eqnarray}
Z^{\prime}=\frac{1}{2}(NSN^{-1}N_{y}SN^{-1}+NSS_{y}N^{-1}-N_{y}N^{-1}+iu^{\prime}NSN^{-1}),\label{3.26} 
\end{eqnarray}
so that we obtain
\begin{eqnarray}
N_{y}N^{-1}=\frac{1}{2}[(NSS_{y}N^{-1}+iuNSN^{-1})-(NSN^{-1}N_{y}SN^{-1}+NSS_{y}N^{-1}-N_{y}N^{-1}+iu^{\prime}NSN^{-1})]\label{3.27} 
\end{eqnarray}
or
\begin{eqnarray}
2N_{y}N^{-1}=NSS_{y}N^{-1}+iuNSN^{-1}-NSN^{-1}N_{y}SN^{-1}-NSS_{y}N^{-1}+N_{y}N^{-1}-iu^{\prime}NSN^{-1}.\label{3.28} 
\end{eqnarray}
So finally we get
\begin{eqnarray}
N_{y}N^{-1}=iuNSN^{-1}-NSN^{-1}N_{y}SN^{-1}-iu^{\prime}NSN^{-1}.\label{3.29} 
\end{eqnarray}
This equation simplies as
\begin{eqnarray}
i(u^{\prime}-u)NSN^{-1}=-(NSN^{-1}N_{y}SN^{-1}+N_{y}N^{-1})\label{3.30} 
\end{eqnarray}
or
\begin{eqnarray}
u^{\prime}-u=i(NSN^{-1}N_{y}N^{-1}+N_{y}SN^{-1})=iN(SN^{-1}N_{y}+N^{-1}N_{y}S)N^{-1}.\label{3.31} 
\end{eqnarray}
Hence we get the DT for the potential $u$:
\begin{eqnarray}
u^{\prime}=u+itr(SN^{-1}N_{y}).\label{3.32} 
\end{eqnarray}

\subsubsection{DT in terms of the $N$ matrix}

We now ready to write the DT for the ML-II equation in the more explicit form. First we collect all DT for the ML-II equation. We have
\begin{eqnarray}
    S^{'}&=& NSN^{-1},  \label{3.33}\\
   u^{'}&=&u+itr(SN^{-1}N_{y}),\label{3.34}\\
    W'&=&(I+\omega N)W(I+\omega N)^{-1}. \label{3.35}
\end{eqnarray}
 It is not difficult to verify that the matrix $N$ has the form
 \begin{equation}
N=\begin{pmatrix} n_{11} & n_{12}\\ -n_{12}^{*} & n_{11}^{*}\end{pmatrix},  \label{3.36}
\end{equation}
so that we have 
\begin{equation} N^{-1}=\frac{1}{n}\begin{pmatrix} n_{11}^{*} & -n_{12}\\ n_{12}^{*} & n_{11}\end{pmatrix},  \label{3.37}
\end{equation}
\begin{equation}
I+\omega N=\begin{pmatrix} n_{11}\omega+1 & \omega n_{12}\\ -\omega n_{12}^{*} & \omega n_{11}^{*}+1\end{pmatrix},\quad 
(I+\omega N)^{-1}=\frac{1}{\square}\begin{pmatrix} \omega n_{11}^{*}+1 & -\omega n_{12}\\ \omega n_{12}^{*} & \omega n_{11}+1\end{pmatrix}. \label{3.38}
\end{equation}
Here
\begin{equation}
n=\det{N}=|n_{11}|^2+|n_{12}|^{2}, \quad \square=det(M+\omega I)=\omega^2(|n_{11}|^{2}+|n_{12}|^{2})+\omega(n_{11}+n_{11}^{*})+1. \label{3.39}
\end{equation}
Finally we have the DT in terms of the elements of $N$ as:
\begin{equation}
S^{'}=\frac{1}{n}\begin{pmatrix}S_{3}(|n_{11}|^{2}-|n_{12}|^{2})+S^{-}n_{11}n_{12}^{*}+S^{+}n^{*}_{11}n_{12} & S^{-}n_{11}^{2}-S^{+}n_{12}^{2}-2S_{3}n_{11}n_{12}\\ S^{+}n_{11}^{*2}-S^{-}n_{12}^{*2}-2S_{3}n_{11}^{*}n_{12}^{*}& S_{3}(|n_{12}|^{2}-|n_{11}|^{2})-S^{-}n_{11}n_{12}^{*}-S^{+}n^{*}_{11}n_{12}\end{pmatrix},\label{3.40}
 \end{equation}
\begin{equation}
u^{'}=
{u+\frac{i}{n}[(n_{11y}n_{11}^{*}+n_{12y}^{*}n_{12}-n_{12y}n_{12}^*-n_{11y}^*n_{11})S_3+(n_{11y}n_{12}^{*}-n_{12y}^*n_{11})S^{-}+(n_{12y}n_{11}^{*}-n_{11y}^*n_{12})S^+]},\label{3.41}
\end{equation}
and
\begin{equation}
W'=\frac{1}{\square}\begin{pmatrix}1+A_{11} & A_{12}\\ A_{21}& -1+A_{22}\end{pmatrix}, \label{3.42}
\end{equation}
where
\begin{equation}
A_{11}=(\omega^2(|n_{11}|^{2}-|n_{12}|^{2})+\omega(n_{11}+n_{11}^{*}))W_3+(\omega n^{*}_{11}+1)\omega n_{12}W^{+}+(\omega n_{11}+1)\omega n^{*}_{12}W^{-}, \label{3.43}
\end{equation}
\begin{equation}
A_{12}={-2\omega^2 n_{11}n_{12}W_3-2\omega n_{12}W_3+\omega^2n_{11}^2W^{-}+2\omega n_{11}W^{-}+W^{-}-\omega^2n_{12}^2W^+},\label{3.44}
\end{equation}
\begin{equation}\label{3.45}
A_{21}=-2\omega^2 n^*_{11}n^*_{12}W_3-2\omega n^*_{12}W_3+\omega^2(n^*_{11})^2W^{+}+2\omega n^{*}_{11}W^{+}+W^{+}-\omega^2(n_{12}^{*})^2W^-,
\end{equation}
\begin{equation}\label{3.46}
A_{22}=-((\omega^2(|n_{11}|^{2}-|n_{12}|^{2})+\omega(n_{11}+n_{11}^{*}))W_3+(\omega n^{*}_{11}+1)\omega n_{12}W^{+}+(\omega n_{11}+1)\omega n^{*}_{12}W^{-}). 
\end{equation}
At last, we give the another form of the DT  of $S$ as:
\begin{eqnarray}
S^{'}=S-iN_{x}=S-i\begin{pmatrix} n_{11x} & n_{12x}\\ -n_{12x}^{*} & n_{11x}^{*}\end{pmatrix}. \label{3.47}
\end{eqnarray}
 \subsubsection{DT in terms of eigenfunctions}
 
 To construct exact solutions of the ML-II equation, we must find the explicit expressions of $n_{ij}$. To do that, we  assume that 
\begin{eqnarray}
N=H\Lambda^{-1} H^{-1}, \label{3.48}
\end{eqnarray}
where \begin{eqnarray}
H=\begin{pmatrix} \psi_{1}(\lambda_{1};t,x,y)&\psi_{1}(\lambda_{2};t,x,y)\\\psi_{2}(\lambda_{1};t,x,y)&\psi_{2}(\lambda_{2};t,x,y)\end{pmatrix}. \label{3.49}
\end{eqnarray}
Here
\begin{eqnarray}
\Lambda&=&\begin{pmatrix} \lambda_{1}&0\\0&\lambda_{2}\end{pmatrix} \label{3.50}
\end{eqnarray}
and $det$ $H\neq0$, where $\lambda_{1}$ and $\lambda_2$ are complex constants.
It is easy to show that $H$ satisfies the following equations
\begin{eqnarray}
H_{x} &=& -iSH\Lambda, \label{3.51}\\
H_{t} &=& 2 H_{y}\Lambda +2ZH\Lambda-\frac{i}{\omega}WH+WH\Sigma, \label{3.52}
\end{eqnarray}
where 
\begin{eqnarray}
Z=0.25([S,S_{y}]+2iuS), \quad \Sigma=\begin{pmatrix} \frac{i}{\lambda_{1}+\omega}&0\\0&\frac{i}{\lambda_{2}+\omega}\end{pmatrix}. \label{3.53}
\end{eqnarray}
  From these equations follow that $N$ obeys the equations
  \begin{eqnarray}
N_{x} &=& iNSN^{-1}-iS, \label{3.54}\\
N_{y} &=& [H_{y}H^{-1},N], \label{3.55}\\
N_{t} &=& 2Z-2Z^{'}-\frac{i}{\omega}(WN-NW)+WH\Sigma\Lambda^{-1}H^{-1}-NWH\Sigma H^{-1}, \label{3.56}
\end{eqnarray}
 These equations  are in fact equivalent to the equations (3.21)-(3.23), respectively as we think. For example, from the equations for $N_{t}$ (3.23) and (3.56) we get
  \begin{eqnarray}
   -\frac{i}{\omega} W'-2Z^{'}+\frac{i}{\omega} NW+2Z=2Z-2Z^{'}-\frac{i}{\omega}(WN-NW)+WH\Sigma\Lambda^{-1}H^{-1}-NWH\Sigma H^{-1}, \label{3.57}
\end{eqnarray}
  or
  \begin{eqnarray}
   -\frac{i}{\omega} W'N=-\frac{i}{\omega}WN+WH\Sigma\Lambda^{-1}H^{-1}-NWH\Sigma H^{-1}. \label{3.58}
\end{eqnarray}
Let us one more rewrite this formula as
\begin{eqnarray}
   -\frac{i}{\omega} W'=-\frac{i}{\omega}W+WH\Sigma H^{-1}-NWH\Sigma \Lambda H^{-1}. \label{3.59}
\end{eqnarray}
  Now we using  the formula  for $W^{'},$ we have
  \begin{eqnarray}
   -\frac{i}{\omega} (I+\omega N)W(I+\omega N)^{-1}=-\frac{i}{\omega}W+WH\Sigma H^{-1}-NWH\Sigma \Lambda H^{-1}. \label{3.60}
\end{eqnarray}
 Hence  we get
  \begin{eqnarray}
   -iWN+\omega WH\Sigma \Lambda^{-1} H^{-1}+WH\Sigma H^{-1}=0. \label{3.61}
\end{eqnarray}
  This equation satisfies identically if we use the following  formulas
   \begin{eqnarray}
   \Sigma \Lambda^{-1}=\frac{i}{\omega}\Lambda^{-1}-\frac{1}{\omega} \Sigma, \quad \Sigma \Lambda=i-\omega  \Sigma. \label{3.62}
\end{eqnarray}
  In order to satisfy the constraints of $S$ and  $W$, the $S$ and the matrix solution of the Lax equations obey the condition
\begin{eqnarray}
\Phi^{\dagger}=\Phi^{-1}, \quad S^{\dagger}=S, \label{3.63}
\end{eqnarray} 
which follow from the equations
\begin{eqnarray}
\Phi^{\dagger}_{x}=i\lambda \Phi^{\dagger}S^{\dagger}, \quad (\Phi^{-1})_{x}=i\lambda \Phi^{-1}S^{-1}, \label{3.64}
\end{eqnarray}
where $\dagger$ denote an Hermitian conjugate. After some calculations we come to the formulas 
\begin{eqnarray}
\lambda_{2}=\lambda^{*}_{1}, \quad
 H=\begin{pmatrix} \psi_{1}(\lambda_{1};t,x,y)&-\psi^{*}_{2}(\lambda_{1};t,x,y)\\ \psi_{2}(\lambda_{1};t,x,y)&\psi^{*}_{1}(\lambda_{1};t,x,y)\\ \end{pmatrix}, \label{3.65}
\end{eqnarray}
\begin{eqnarray}
H^{-1}=\frac{1}{\Delta}\begin{pmatrix} \psi^{*}_{1}(\lambda_{1};t,x,y)&\psi^{*}_{2}(\lambda_{1};t,x,y)\\ -\psi_{2}(\lambda_{1};t,x,y)&\psi_{1}(\lambda_{1};t,x,y)\\ \end{pmatrix}, \label{3.66}
\end{eqnarray}
where 
\begin{eqnarray}
\Delta &=&|\psi_{1}|^2+|\psi_{2}|^2. \label{3.67}
\end{eqnarray}
So for the matrix $N$ we have
\begin{eqnarray}
N&=&\frac{1}{\Delta}\begin{pmatrix} \lambda_{1}^{-1}|\psi_{1}|^2+\lambda^{-1}_{2}|\psi_{2}|^2 & (\lambda_{1}^{-1}-\lambda_{2}^{-1})\psi_{1}\psi_{2}^{*}\\ (\lambda_{1}^{-1}-\lambda_{2}^{-1})\psi_{1}^{*}\psi_{2} & \lambda_{1}^{-1}|\psi_{2}|^2+\lambda_{2}^{-1}|\psi_{1}|^2)\end{pmatrix}.  \label{3.68}
\end{eqnarray}
 Now  let us rewrite the 1-fold DT in the more unified form:
\begin{equation}
\Phi^{[1]}=L_1\Phi, \label{3.69}
\end{equation}
where
\begin{equation}
L_{1}=\lambda l^{1}_{1}+l_{1}^{0}=\lambda l^{1}_{1}-I.\label{3.70}
\end{equation}
Then the 1-fold DT can be written as
\begin{eqnarray}
    S^{[1]}&=& l^{1}_{1}S(l^{1}_{1})^{-1},  \label{3.71}\\
   u^{[1]}&=&u+itr[S(l^{1}_{1})^{-1}l^{1}_{1y}],\label{3.72}\\
    W^{[1]}&=&L_{1}|_{\lambda=-\omega}WL_{1}^{-1}|_{\lambda=-\omega}. \label{3.73}
\end{eqnarray}

\subsection{n-fold DT}

As the 1-fold DT, now we can construct the n-fold DT. In this case we have the following transformation for eigenfunctions:
\begin{equation}
\Phi^{[n]}=L_n\Phi^{[n-1]}=(\lambda N^{[n]}-I)\Phi^{[n-1]}=(\lambda N_n-I)\ldots(\lambda N_2-I)(\lambda N_1-I)\Phi \label{3.74}
\end{equation}
so that
\begin{equation}
\Phi^{[n]}=[\lambda^n l_{n}^n+\lambda^{n-1} l_{n}^{n-1}+ ... +\lambda l^{1}_{n}+l_{n}^{0}]\Phi. \label{3.75}
\end{equation}

For the $n$-fold DT of the ML-II equation the matrix function $L_{n}$ satisfies the equations
\begin{eqnarray}
L_{nx}&=&U^{[n]}L_{n}-L_{n}U, \label{3.76}\\
L_{nt}&=&2\lambda L_{ny}+V^{[n]}L_{n}-T_{n}V. \label{3.77}
\end{eqnarray}
As result,  we obtain 
\begin{eqnarray}
U^{[n]}&=&L_{nx}L_{n}^{-1}+L_{n}UL_{n}^{-1}, \label{3.78}\\
V^{[n]}&=&L_{nt}L_{n}^{-1}-2\lambda L_{ny}L_{n}^{-1}+L_{n}VL_{n}^{-1}. \label{3.79}
\end{eqnarray}

\section{Soliton solutions}

In this section, we apply the above constructed DT to find the 1-soliton and 2-soliton solutions of the ML-II equation.  
\subsection{1-soliton solution}
 To find the 1-soliton solution, we consider the following  seed solution 
 \begin{eqnarray}
    S=\sigma_{3}, \quad u=0, \quad W=b\sigma_{3}, \label{4.1}
\end{eqnarray}where $b=const$. Then we get
\begin{eqnarray}
    S^{'}&=& \frac{1}{n}\begin{pmatrix}|n_{11}|^{2}-|n_{12}|^{2} & -2n_{11}n_{12}\\ -2n_{11}^{*}n_{12}^{*}& |n_{12}|^{2}-|n_{11}|^{2})\end{pmatrix}, \label{4.2}\\
   u^{'}&=&\frac{i}{n}(n_{11y}n_{11}^*+n_{12y}^{*}n_{12}-n_{12y}n_{12}^*-n_{11y}^*n_{11})\label{4.3}
  \end{eqnarray} 
  and
   \begin{equation}
    W'=\frac{1}{\square}\begin{pmatrix}{b(\omega^2(|n_{11}|^{2}-n_{12}|^{2})+\omega(n_{11}+n_{11}^{*})+1}) & -2\omega n_{11}n_{12}-2n_{12})\\ 
    -2\omega n^{*}_{11}n^{*}_{12}-2n^{*}_{12}& -b(\omega^2(|n_{11}|^{2}-n_{12}|^{2})+\omega(n_{11}+n_{11}^{*})+1)\end{pmatrix}. \label{4.4}
\end{equation}
Now we  are ready to write the solutions of the ML-II in terms of the elements of $N$. We get
\begin{eqnarray}
    S^{+\prime}&=&-\frac{2n_{11}^{*}n_{12}^{*}}{n} ,  \label{4.5}\\
   S^{-\prime}&=&-\frac{2n_{11}n_{12}}{n},\label{4.6}\\
    S^{\prime}_{3}&=&\frac{|n_{11}|^{2}-|n_{12}|^{2}}{n}, \label{4.7}\\
    W^{+\prime}&=&\frac{-2b\omega n^*_{12}(\omega n^*_{11}+1)}{ \square}, \label{4.8} \\
   W^{-\prime}&=&\frac{-2b\omega n_{12}(\omega n_{11}+1)}{\square},\label{4.9}\\
    W^{\prime}_{3}&=&\frac{\omega^2(|n_{11}|^{2}-|n_{12}|^{2})+\omega(n_{11}+n_{11}^{*})+1)b}{\square}. \label{4.10}\\
    u^{'}&=&\frac{i}{n}(n_{11y}n_{11}^*+n_{12y}^{*}n_{12}-n_{12y}n_{12}^*-n_{11y}^*n_{11}).\label{4.11}
    \end{eqnarray}
    To get the 1-soliton solution we need the explicit expressions for eigenfunctions. They come from the system \begin{eqnarray}
\psi_{1x}&=&-i\lambda \psi_1,\label{4.12}\\
\psi_{2x}&=&i\lambda \psi_2,\label{4.13}\\
\psi_{1t}&=&2\lambda\psi_{1y}+ib(\frac{1}{\lambda+\omega}-\frac{1}{\omega})\psi_1,\label{4.14} \\
\psi_{2t}&=&2\lambda\psi_{2y}-ib(\frac{1}{\lambda+\omega}-\frac{1}{\omega})\psi_2\label{4.15} 
\end{eqnarray}  
and have the form
\begin{eqnarray}
\psi_{1}&=&e^{\theta_1+i\chi_1},\label{4.16}\\
\psi_{2}&=&e^{\theta_2+i\chi_2}.\label{4.17}
\end{eqnarray}
Here $\lambda=\alpha+i\beta$, $b_{i}=\mu_i+i\nu_i$ and $\delta_i=\tau_i+i\sigma_i$ and  $\alpha,\beta,\mu,\nu,\tau,\sigma,\delta_0$ are real constants, $\theta_1=-\theta_2,\quad \chi_2=-\chi_1+\delta_0$ with
  \begin{eqnarray}
 \theta_1&=&\beta x-\nu_1y-(2\alpha\nu_1+2\beta\mu_1-\frac{\beta b}{M})t-\sigma_1,\label{4.18}\\
 \theta_2&=&-\beta x-\nu_2y-(2\alpha\nu_2+2\beta\mu_2+\frac{\beta b}{M})t-\sigma_2,\label{4.19}\\
 \chi_1&=&-\alpha x+\mu_1y+\left[2\alpha\mu_1-2\beta\nu_1+\frac{b(\alpha+\omega)}{M}-\frac{b}{\omega}\right]t+\tau_1,\label{4.20}\\
 \chi_2&=&\alpha x+\mu_2y+\left[2\alpha\mu_2-2\beta\nu_2-\frac{b(\alpha+\omega)}{M}+\frac{b}{\omega}\right]t+\tau_2+\delta_{0}. \label{4.21}  
 \end{eqnarray}
So for the elements of $N$ we obtain the following expressions 
 \begin{eqnarray}
  n_{11}&=&\frac{1}{\alpha^2+\beta^2}(\alpha-i\beta\tanh2\theta_1),\label{4.22}\\
  n_{12}&=&\frac{-i\beta e^{2i\chi_1-i\delta_0}}{(\alpha^2+\beta^2)\cosh2\theta_1}.\label{4.23}
\end{eqnarray}
 Now we can write the 1-soliton solution of the ML-II equation as
\begin{eqnarray} S^{[1]}_{3}&=&\tanh^22\theta_1+\frac{\alpha^2-\beta^2}{\alpha^2+\beta^2}\frac{1}{\cosh^22\theta_1}, \label{4.24}\\
S^{+[1]}&=&\frac{2\beta}{\alpha^2+\beta^2}e^{-i\chi_1+i\chi_2}\left(\frac{\beta\sinh2\theta_1}{\cosh^22\theta_1}-\frac{i\alpha}{\cosh 2\theta_1}\right), \label{4.25}\\
 S^{-[1]}&=&\frac{2\beta}{\alpha^2+\beta^2}e^{i\chi_1-i\chi_2}\left(\frac{\beta\sinh 2\theta_1}{\cosh^22\theta_1}+\frac{i\alpha}{\cosh2\theta_1}\right),\label{4.26}\\
u^{[1]}&=&\frac{4\beta(\beta\mu_{1}-\alpha\nu_{1})}{(\alpha^2+\beta^2)\cosh^{2} 2\theta_1}\label{4.27}
\end{eqnarray}
 and
\begin{eqnarray}
W^{[1]}_{3}&=&\frac{({\frac{\omega^2}{\alpha^2+\beta^2}S^{\prime}_{3}+\frac{2\alpha\omega}{\alpha^2+\beta^2}+1})b}{{\frac{\omega^2}{\alpha^2+\beta^2}+\frac{2\alpha\omega}{\alpha^2+\beta^2}+1}}=\left(\omega^2S^{\prime}_{3}+2\alpha\omega+\frac{1}{\alpha^2+\beta^2}\right)\frac{b}{M}, \label{4.28}\\
W^{+[1]}&=&\frac{\left({\frac{\omega^2}{\alpha^2+\beta^2}S^{+\prime}-\frac{2i\beta\omega e^{-2i\chi_1+i\delta_0}}{(\alpha^2+\beta^2)\cosh2\theta_1}}\right)b}{{\frac{\omega^2}{\alpha^2+\beta^2}+\frac{2\alpha\omega}{\alpha^2+\beta^2}+1}}=\left(\omega^2S^{+\prime}-\frac{2i\beta\omega e^{-2i\chi_1+i\delta_0}}{\cosh2\theta_1}\right)\frac{b}{M}, \label{4.29}\\
W^{-[1]}&=&\frac{\left({\frac{\omega^2}{\alpha^2+\beta^2}S^{-\prime}+\frac{2i\beta\omega e^{2i\chi_1-i\delta_0}}{(\alpha^2+\beta^2)\cosh2\theta_1}}\right)b}{{\frac{\omega^2}{\alpha^2+\beta^2}+\frac{2\alpha\omega}{\alpha^2+\beta^2}+1}}=\left(\omega^2S^{-\prime}+\frac{2i\beta\omega e^{2i\chi_1-i\delta_0}}{\cosh2\theta_1}\right)\frac{b}{M},\label{4.30}
\end{eqnarray}
where $M=(\omega+\alpha)^2+\beta^2.$
The expressions of $W$ we can rewrite  as
\begin{eqnarray}
W^{[1]}_{3}&=&\left(\frac{\omega^2}{\cosh^22\theta_1}\left(\sinh^22\theta_1+\frac{\alpha^2-\beta^2}{\alpha^2+\beta^2}\right)+2\alpha\omega+\frac{1}{\alpha^2+\beta^2}\right)\frac{b}{M}, \label{4.31}\\
 W^{+[1]}&=&\frac{2\beta\omega e^{-2i\chi_1+i\delta_0}b}{(\alpha^2+\beta^2)M\cosh2\theta_1}\left(\beta\omega\tanh 2\theta_1-i(\alpha\omega+\alpha^2+\beta^2)\right), \label{4.32}\\
  W^{-[1]}&=&\frac{2\beta\omega e^{2i\chi_1-i\delta_0}b}{(\alpha^2+\beta^2)M\cosh2\theta_1}\left(\beta\omega\tanh 2\theta_1+i(\alpha\omega+\alpha^2+\beta^2)\right).\label{4.33}
\end{eqnarray}
Also we note that
\begin{eqnarray}
W^{[1]}=\frac{\omega^2b}{M}S^{[1]}+F^{[1]}, \label{4.34}
\end{eqnarray}
where
 \begin{eqnarray}
F^{[1]}_{3}&=&\frac{b}{M}\left(2\alpha\omega+\frac{1}{\alpha^2+\beta^2}\right), \label{4.35}\\
F^{+[1]}&=&-\frac{2i\beta\omega b e^{-2i\chi_1-i\delta_0}}{M\cosh2\theta_1}, \label{4.36}\\
F^{-[1]}&=&\frac{2i\beta\omega b e^{2i\chi_1+i\delta_0}}{M\cosh2\theta_1}.\label{4.37}
\end{eqnarray}

Now let us briefly consider some particular cases. 

i) First we assome that $\lambda_{1}=i\beta$ that is $\alpha=0$. In this case we have
\begin{eqnarray}
  n_{11}&=&-\frac{i}{\beta}\tanh2\theta_1^{\prime},\label{4.38}\\
  n_{12}&=&\frac{-ie^{2i\chi_1^{\prime}-i\delta_0}}{\beta\cosh2\theta_1^{\prime}},\label{4.39}
\end{eqnarray}
where $\theta_1^{\prime}=\theta_1|_{\alpha=0}$ and $\chi_1^{\prime}=\chi_1|_{\alpha=0}$. The corresponding 1-soliton solution takes the form
\begin{eqnarray}
S^{[1]}_{3}&=&\tanh^22\theta_1^{\prime}-\frac{1}{\cosh^22\theta_1^{\prime}}, \label{4.40}\\
S^{+[1]}&=&\frac{2e^{-i\chi_1^{\prime}+i\chi_2^{\prime}}\sinh2\theta_1^{\prime}}{\cosh^22\theta_1^{\prime}}, \label{4.41}\\
 S^{-[1]}&=&\frac{2e^{i\chi_1^{\prime}-i\chi_2^{\prime}}\sinh2\theta_1^{\prime}}{\cosh^22\theta_1^{\prime}},\label{4.42}\\
u^{[1]}&=&\frac{4\mu_{1}}{\cosh^{2} 2\theta_1^{\prime}},\label{4.43}
\end{eqnarray}
 and
 \begin{eqnarray}
W^{[1]}_{3}&=&\left[\frac{\omega^2}{\cosh^22\theta_1^{\prime}}\left(\sinh^22\theta_1^{\prime}-1\right)+\frac{1}{\beta^2}\right]\frac{b}{M^{\prime}}, \label{4.44}\\
 W^{+[1]}&=&\frac{2\omega e^{-2i\chi_1^{\prime}+i\delta_0}b}{ M^{\prime}\cosh2\theta_1^{\prime}}\left(\omega\tanh 2\theta_1^{\prime}-i\beta\right), \label{4.45}\\
  W^{-[1]}&=&\frac{2\omega e^{2i\chi_1^{\prime}-i\delta_0}b}{M^{\prime}\cosh2\theta_1^{\prime}}\left(\omega\tanh 2\theta_1^{\prime}+i\beta\right).\label{4.46}
\end{eqnarray}
where $M^{\prime}=\omega^2+\beta^2.$

ii) Now we consider the case $\lambda_{1}=\alpha$ that $\beta=0$. Then we have
\begin{eqnarray}
  n_{11}&=&\frac{1}{\alpha},\label{4.47}\\
  n_{12}&=&0. \label{4.48}
\end{eqnarray}
 Now we can write the 1-soliton solution of the ML-II equation as
\begin{eqnarray}
S^{[1]}_{3}&=&1, \label{4.49}\\
 S^{+[1]}&=&0, \label{4.50}\\
 S^{-[1]}&=&0,\label{4.51}\\
u^{[1]}&=&0\label{4.52}
\end{eqnarray}
 and
\begin{eqnarray}
W^{[1]}_{3}&=&\left(\omega^2+2\alpha\omega+\frac{1}{\alpha^2}\right)\frac{b}{(\omega+\alpha)^2}, \label{4.53}\\
W^{+[1]}&=&0, \label{4.54}\\
W^{-[1]}&=&0. \label{4.55}
\end{eqnarray}

\subsection{2-soliton solution}

Now we are  going to construct the 2-soliton solution of the ML-II equation. For this purpose, let us consider the following transformation 
\begin{eqnarray}
\Phi^{[2]}=L_{2}\Phi^{[1]}, \label{4.56}
\end{eqnarray}
where we assume that
\begin{eqnarray}
L_{2}=\lambda N^{[2]}-I \label{4.57}
\end{eqnarray}
with  
\begin{eqnarray}
N^{[2]}=\begin{pmatrix} n_{11}^{[1]}&n_{12}^{[1]}\\-n_{12}^{[1]*}&n_{11}^{[1]*}\end{pmatrix}. \label{4.58}
\end{eqnarray}
Here  $\Phi^{[j]}$ satisfy the systems
\begin{eqnarray}
\Phi^{[2]}_{x} &=& U^{[2]}\Phi^{[2]},\label{4.59}\\
\Phi^{[2]}_{t} &=& 2\lambda\Phi^{[2]}_{y}+V^{[2]}\Phi^{[2]} \label{4.60}
\end{eqnarray}
and
\begin{eqnarray}
\Phi^{[1]}_{x} &=& U^{[1]}\Phi^{[1]},\label{4.61}\\
\Phi^{[1]}_{t} &=& 2\lambda\Phi^{[1]}_{y}+V^{[1]}\Phi^{[1]}. \label{4.62}
\end{eqnarray}
The  matrix  $L_{[2]}$ obeys the following equations
\begin{eqnarray}
L_{2x}+L_{2}U^{[1]} &=& U^{[2]}L_{2},\label{4.63} \\
L_{2t}+L_{2}V^{[1]} &=& 2\lambda L_{2y}+V^{[2]}L_{2}. \label{4.64}
\end{eqnarray}
In terms of eigenfunctions,  the matrix $N^{[2]}$ takes the form 
\begin{eqnarray}
N^{[2]}&=&\frac{1}{\Delta^{[1]}}\begin{pmatrix} \lambda_{3}^{-1}|\psi_{1}^{[1]}|^2+\lambda^{-1}_{4}|\psi_{2}^{[1]}|^2 & (\lambda_{3}^{-1}-\lambda_{4}^{-1})\psi_{1}^{[1]}\psi_{2}^{[1]*}\\ (\lambda_{3}^{-1}-\lambda_{4}^{-1})\psi_{1}^{[1]*}\psi_{2}^{[1]} & \lambda_{3}^{-1}|\psi_{2}^{[1]}|^2+\lambda_{4}^{-1}|\psi_{1}^{[1]}|^2)\end{pmatrix},  \label{4.65}
\end{eqnarray}
where $\lambda_{4}=\lambda_{3}^{*}$. To find  elements of $\Phi^{[1]}$ we use the one-fold DT saying the formula
\begin{eqnarray}
\Phi^{[1]}=L_{1}\Phi^{[0]}. \label{4.66}
\end{eqnarray}
Here
\begin{eqnarray}
\Phi^{[0]}&=&\begin{pmatrix}\psi_{1}(\lambda_{1}) & -\psi_{2}^{*}(\lambda_{1})\\ \psi_{2}(\lambda_{1}) & \psi_{1}^{*}(\lambda_{1})\end{pmatrix}  \label{4.67}
\end{eqnarray}
and
\begin{eqnarray}
L_{1}=\begin{pmatrix}\lambda n_{11}(\lambda_{1})-1 & \lambda n_{12}(\lambda_{1})\\-\lambda n_{12}^{*}(\lambda_{1})& \lambda n_{11}^{*}(\lambda_{1})-1\end{pmatrix}.  \label{4.68}
\end{eqnarray}
So we finally have
\begin{eqnarray}
\Phi^{[1]}=\begin{pmatrix}(\lambda n_{11}-1)\psi_{1}+ \lambda n_{12}\psi_{2} & -(\lambda n_{11}-1)\psi_{2}^{*}+ \lambda n_{12}\psi_{1}^{*}\\ -\lambda n_{12}^{*} \psi_{1}+(\lambda n_{11}^{*}-1)\psi_{2}&\lambda n_{12}^{*} \psi_{2}^{*}+(\lambda n_{11}^{*}-1)\psi_{1}^{*}\end{pmatrix}=\begin{pmatrix} \psi_{1}^{[1]} &-\psi_{2}^{[1]*}\\ \psi_{2}^{[1]} & \psi_{1}^{[1]*}\end{pmatrix}. \label{4.69}
\end{eqnarray}
Here
\begin{eqnarray}
\psi^{[1]}_{1}&=&\frac{(\alpha_3+i\beta_3)e^{i\chi_1}}{\alpha^{2}_1+\beta^{2}_1}\left[\left(\alpha_1-i\beta_1\tanh 2\theta_1-\frac{\alpha^{2}_1+\beta^{2}_1}{\alpha_3+i\beta_3}\right)e^{\theta_1}-\frac{i\beta_1e^{-\theta_1}}{\cosh 2\theta_1}\right]\label{4.70},\\
\psi^{[1]}_{2}&=&\frac{(\alpha_3+i\beta_3)e^{-i\chi_1+i\delta_0}}{\alpha^{2}_1+\beta^{2}_1}\left[\left(\alpha_1+i\beta_1\tanh 2\theta_1-\frac{\alpha^{2}_1+\beta^{2}_1}{\alpha_3+i\beta_3}\right)e^{-\theta_1}-\frac{i\beta_1e^{\theta_1}}{\cosh 2\theta_1}\right]\label{4.71},\\
\Delta^{[1]}&=&\frac{2(\alpha_3+i\beta_3)^2\cosh2\theta_1}{(\alpha^2_1+\beta^2_1)^2}\left(A+\frac{\beta^2_1}{\cosh^22\theta_1}\right).\label{4.72}
\end{eqnarray}

The 2-soliton solution of the ML-II equation has the form
 \begin{eqnarray}
S^{[2]}_{3}&=&\frac{1}{n^{[1]}}\left[S_{3}^{[1]}(|n_{11}^{[1]}|^{2}-|n_{12}^{[1]}|^{2})+S^{-[1]}n_{11}^{[1]}n_{12}^{[1]*}+S^{+[1]}n^{[1]*}_{11}n_{12}^{[1]}\right], \\
S^{-[2]}&=&\frac{1}{n^{[1]}}\left[S^{-[1]}n_{11}^{[1]2}-S^{+[1]}n_{12}^{[1]2}-2S_{3}^{[1]}n_{11}^{[1]}n_{12}^{[1]}\right], \label{4.77}\\
S^{+[2]}&=&\frac{1}{n^{[1]}}\left[S^{+[1]}n_{11}^{[1]*2}-S^{-[1]}n_{12}^{[1]*2}-2S_{3}^{[1]}n_{11}^{[1]*}n_{12}^{[1]*}\right], \label{4.77}\\
 u^{[2]}&=&
u^{[1]}+\frac{i}{n^{[1]}}[(n_{11y}^{[1]}n_{11}^{[1]*}+n_{12y}^{[1]*}n_{12}^{[1]}-n_{12y}^{[1]}n_{12}^{[1]*}-n_{11y}^{[1]*}n_{11}^{[1]})S_3^{[1]}+\nonumber \\
& & (n_{11y}^{[1]}n_{12}^{[1]*}-n_{12y}^{[1]*}n_{11}^{[1]})S^{-[1]}+(n_{12y}^{[1]}n_{11}^{[1]*}-n_{11y}^{[1]*}n_{12}^{[1]})S^{+[1]}], \label{4.78}\\
W^{[2]}_{3}&=&\frac{1+A_{11}^{[2]}}{\square}, \label{}\\
W^{-[2]}&=&\frac{A_{12}^{[2]}}{\square}, \label{}\\
W^{+[2]}_{3}&=&\frac{A_{21}^{[2]}}{\square}, \label{}
\end{eqnarray}
where
\begin{equation}\tiny
A_{11}^{[2]}=\left[\omega^2(|n_{11}^{[1]}|^{2}-|n_{12}^{[1]}|^{2})+\omega(n_{11}^{[1]}+n_{11}^{[1]*})\right]W_3^{[1]}+(\omega n^{[1]*}_{11}+1)\omega n_{12}^{[1]}W^{+[1]}+(\omega n_{11}^{[1]}+1)\omega n^{[1]*}_{12}W^{-[1]}, \label{4.80}
\end{equation}
\begin{equation}\tiny 
A_{12}^{[2]}=-2\omega^2 n_{11}^{[1]}n_{12}^{[1]}W_3^{[1]}-2\omega n_{12}^{[1]}W_3^{[1]}+(\omega^2n_{11}^{[1]2}+2\omega n_{11}^{[1]}+1)W^{-[1]}-\omega^2n_{12}^{[1]2}W^{+[1]},\label{4.81}
\end{equation}
\begin{equation}\tiny
A_{21}^{[2]}=-2\omega^2 n^{[1]*}_{11}n^{[1]*}_{12}W_3^{[1]}-2\omega n^{[1]*}_{12}W_3^{[1]}+(\omega^2(n^{[1]*}_{11})^2+2\omega n^{[1]*}_{11}+1)W^{+[1]}-\omega^2(n_{12}^{[1]*})^2W^{-[1]},\label{4.82}
\end{equation}
\begin{equation}\tiny
A_{22}^{[2]}=-\left[(\omega^2(|n_{11}^{[1]}|^{2}-|n_{12}^{[1]}|^{2})+\omega(n_{11}^{[1]}+n_{11}^{[1]*})\right]W_3^{[1]}+(\omega n^{[1]*}_{11}+1)\omega n_{12}^{[1]}W^{+[1]}+(\omega n_{11}^{[1]}+1)\omega n^{[1]*}_{12}W^{-[1]}. \label{4.83}
\end{equation}
Here $S^{+[1]}, S^{-[1]}, S^{[1]}_{3}, u^{[1]}, W^{+[1]}, W^{-[1]}, W_{3}^{[1]}$ are given by (\ref{4.24})-(\ref{4.30}), $\square=\omega^2(|n_{11}^{[1]}|^{2}+|n_{12}^{[1]}|^{2})+\omega(n_{11}^{[1]}+n_{11}^{[1]*})+1$ and
\begin{eqnarray}
n_{11}^{[1]}&=&\frac{1}{\alpha_3^2+\beta^2_3}\left[\alpha_3-i\beta_3\left(\frac{A\cosh^22\theta_1-\beta^2_1}{A\cosh^2 2\theta_1+\beta^2_1}\right)\tanh 2\theta_1\right],\label{4.73}\\
  n_{12}^{[1]}&=&-\frac{i\beta_3e^{2i\chi_1-i\delta_0}}{(\alpha^{2}_3+\beta^{2}_3)\cosh 2\theta_1}\left[\frac{B\cosh^22\theta_1 +\beta^2_1}{A\cosh^22\theta_1+\beta^2_1}\right],\label{4.74}
\end{eqnarray}
where
\begin{equation}
A=\alpha^{2}_1+\beta^2_1\tanh^22\theta_1-\frac{2(\alpha_1\alpha_3-\beta_1\beta_3\tanh2\theta_1)}{\alpha^{2}_3+\beta^{2}_3}+\frac{(\alpha^{2}_1+\beta^{2}_1)^2}{\alpha^{2}_3+\beta^{2}_3}, \label{4.75}
\end{equation}
\begin{equation}
B=\alpha_1^2-2i\beta_1\tanh 2\theta_1-\beta^2_1\tanh^22\theta_1+\frac{2i(\alpha^{2}_1+\beta^{2}_1)\alpha_3\beta_1\tanh2\theta_1}{\alpha^{2}_3+\beta^{2}_3}+\frac{(\alpha^{2}_1+\beta^{2}_1)^2}{(\alpha^{2}_3+\beta^{2}_3)}.\label{4.76}
\end{equation}

 \section{Conclusion}
 
 In this paper, we have considered the integrable ML-II equation which is one of  (2+1)-dimensional generalizations of Heisenberg ferromagnetic equation with self-consistent potentials. We have derived the DT for the ML-II equation.  Using DT and seed solutions, we have constructed the 1-soliton solution and 2-soliton solution.   Similarly, using the higher order DT, one can also generate multi-soliton solutions as well as other types exact solutions like positon, rogue wave, breather etc.  There are still some questions which are still open. One of main questions is the physical interpretations and observation of the above constructed soliton solutions in magnetic physics. We hope at least some of these open questions will closed in future investigations.  Another interesting question is the dynamical interactions of solitons which also can be analysed in the DT.

 \end{document}